\documentclass[aip,jcp,amsmath,amssymb,reprint,floatfix]{revtex4-2}

\usepackage{graphicx}
\usepackage{dcolumn}
\usepackage{bm}
\usepackage[utf8]{inputenc}
\usepackage[T1]{fontenc}
\usepackage{mathptmx}
\usepackage{booktabs}
\usepackage{color}
\usepackage{multirow}

\usepackage{notes2bib}
\begin{document}

\title{Assessing the Global Natural Orbital Functional Approximation on Model Systems with Strong Correlation}

\author{Ion Mitxelena}
\email{ion.mitxelena@unirioja.es}
\affiliation{Área de Física, Departamento de Química, Universidad de la Rioja, 26006 Logroño, La Rioja, Spain}

\author{Mario Piris}
\email{mario.piris@ehu.eus}
\affiliation{Donostia International Physics Center (DIPC) \& Euskal Herriko Unibertsitatea (UPV/EHU), 20018 Donostia, Euskadi, Spain}
\affiliation{Basque Foundation for Science (IKERBASQUE), 48013 Bilbao, Euskadi, Spain}

\date{\today}

\begin{abstract}
In the past decade, natural orbital functional (NOF) approximations have emerged as prominent tools for characterizing electron correlation. Despite their effectiveness, these approaches, which rely on natural orbitals and their associated occupation numbers, often require hybridization with other methods to fully account for all correlation effects. Recently, a global NOF (GNOF) has been proposed [Phys. Rev. Lett. 127, 233001 (2021)] to comprehensively address both dynamic and static correlations. This study evaluates the performance of GNOF on strongly correlated model systems, including comparisons with highly accurate Full Configuration Interaction (FCI) calculations for hydrogen atom clusters in one, two, and three dimensions. Additionally, the investigation extends to a BeH$_{2}$ reaction, involving the insertion of a beryllium atom into a hydrogen molecule along a C$_{2v}$ pathway. According to the results obtained using GNOF, consistent behavior is observed across various correlation regions, encompassing a range of occupation and orbital schemes. Furthermore, distinctive features are identified when varying the dimensionality of the system.
\end{abstract}

\maketitle

\section{Introduction}

Since the electron-electron interaction is described by a two-body operator in the non-relativistic Born-Oppenheimer approximation, the resulting energy is an exact functional of the second-order reduced density matrix (2RDM). Although the existence of the functional in terms of the first-order RDM (1RDM) was proven \cite{Valone1980} and remains the subject of research \cite{SP21,CLLPPS23}, its specific form remains unknown. The adopted strategy for electronic calculations involves the reconstruction of the 2RDM in terms of the 1RDM. However, the energy obtained through this process is not the exact functional of the 1RDM. Consequently, an approximate functional in this manner retains a hidden dependence on the 2RDM, leading to the consequential functional N-representability problem \cite{Ludena2013,Piris2018-nrep}. The question arises: How can we express the 2RDM in terms of the 1RDM?

In 1956, Löwdin and Shull demonstrated \cite{lowdin:56pr} that by constraining the 1RDM to the diagonal representation, we can express the two-electron wavefunction in the simplest manner. In this representation, the diagonal elements of the 1RDM and their corresponding eigenvectors are denoted as occupation numbers (ONs) and natural orbitals (NOs), respectively. The most straightforward generalization of two-electron functional form to N-particle systems is to divide the system into independent electron pairs that leads to the Piris natural orbital functional 5 (PNOF5) \cite{piris2011-pnof5}. An interesting relationship between geminal approaches and NOF theory emerges here, as PNOF5 has been demonstrated to be equivalent to a specific case of an antisymmetrized product of strongly orthogonal geminals (APSG) \cite{Pernal2013,Piris2013e}. Additionaly, PNOF7 \cite{Piris2017,mitxelena2018a} mantains PNOF5 intra-pair interactions but includes inter-pair correlation with an antisymmetrized geminal power (AGP) like term.

Despite recent advances obtained in the context of geminal wavefunction approaches \cite{gaikwad-jcp} and Richardson-Gaudin states \cite{johnson-deprince}, including dynamic correlation effects continue being a challenge. The same arguments motivated the development of a perturbative correction to PNOF5 \cite{Piris2013c}. In this line, hybrid approaches have been recently extensively explored in NOF theory \cite{hollet-titou-2020,mauricio-2021-jctc,wenna-jpcl-2022,elayan-hollet-2022,mazziotti-PRL-2023-NOFT}. For instance, Wenna \textit{et al.} \cite{wenna-jpcl-2022} combined DFT and NOF with the aim to exploit the former in the equilibrium region and the latter for long-range interactions. More recently, Gibney and co-workers \cite{mazziotti-PRL-2023-NOFT} introduced a damping factor to avoid double counting in a functional with terms from both DFT and NOF approximations. Nevertheless, the access to correlated density matrices is limited in this context and thereby it is preferable to have a fully correlated energy functional. A global NOF (GNOF) has been recently proposed \cite{piris-prl-2021} as a method to deal with global electronic structure problems. Despite being an electron-pairing approach, GNOF is not an independent-pair approximation anymore. This functional introduces interactions between electron pairs to treat in a balanced manner the static and dynamic electron correlations \cite{mitxelena_2022,felipe-jctc-2023-porphyrin,felipe-2023-jcp-delocalization,Franco2023,mercero-txema-2023-aromaticity}. Therefore, in contrast to its predecessors \cite{Piris2013c,Piris2017,piris-2018-dyn,mitxelena2018a}, GNOF does not require perturbative corrections to include dynamic correlation.

The motivation that brings us to this study is the assessment of GNOF. Although NOF approximations have been proposed for two decades now, benchmarking new functionals is essential to establish them among daily used methods. The set of systems chosen for this purpose has recently been employed by Gaikwad and co-workers \cite{gaikwad-jcp}. In our previous study \cite{mitxelena_2022}, few of these systems demonstrated the capability of GNOF to accurately describe H-model systems with strong correlation, which can be computed at the Full Configuration Interaction (FCI) level, rendering them suitable for comparisons.  A more complete and challenge set is included in the present work, where we focus on the performance of the GNOF approximation on energy curves. 

Note that throughout the article, strong correlation is used to refer to systems characterized by near degeneracy of the NOs and corresponding ONs, where static electron correlation becomes important. In these situations, there are ONs that differ substantially from zero and one, taking values close to half, making them impossible to describe by a single Slater determinant.

This article is structured as follows: The first section (\ref{sec:theory}) emphasizes the key aspects of the GNOF approximation for spin-compensated systems. The subsequent section (\ref{sec:results}) analyzes a set of model systems, specifically conducting a comparison between NOF calculations and FCI. Finally, conclusions are drawn in section \ref{sec:conclusions}.

\section{Natural Orbital Functional Approximations}\label{sec:theory}

In this section, we provide a brief description of GNOF for singlet systems. A comprehensive review of recent developments in approximate NOFs has just been published \cite{piris2023advances}. GNOF was proposed as a versatile method for addressing various electronic structure problems \cite{piris-prl-2021}. It has demonstrated accurate comparisons with experiments and state-of-the-art calculations for energy differences between ground states and low-lying excited states with different spin states of atoms ranging from H to Ne, ionization potentials of first-row transition-metal atoms (Sc-Zn), total energies of a selected set of 55 molecular systems in different spin states, homolytic dissociation of selected diatomic molecules in different spin states, and the rotation barrier of ethylene. More recently, GNOF has been successfully employed to reduce the delocalization error in NOF approximations \cite{felipe-2023-jcp-delocalization}, conduct \textit{ab-initio} molecular dynamics simulations \cite{alejandro-piris-dynamics-2023}, develop excited state methods by coupling GNOF with extended random phase approximation \cite{felipe-2023-erpa}, and, most importantly for our purposes, investigate problems with high static correlation such as the aromaticity of the Al$_3^{-}$ anion \cite{mercero-txema-2023-aromaticity} and Iron(II) Porphyrin \cite{felipe-jctc-2023-porphyrin}, which poses a challenge due to its computational demands.

The most common strategy for developing NOFs is to approximate the 2RDM in terms of the ONs and NOs. This reconstruction is not applicable in another representation of the 1RDM matrix but in its diagonal form. Therefore, it is more precise to refer to NOF theory rather than 1RDM theory. The exact non-relativistic energy functional of the 2RDM is employed, leading to the solution being established by optimizing the energy with respect to the ONs and the NOs separately. As a result, orbitals vary throughout the optimization process until the most favorable orbital interactions are attained.

Hereafter we focus on spin compensated systems ($S=0$). Consequently, the spin-restricted theory can be applied, ensuring that all spatial orbitals are doubly occupied with equal occupancies for particles with $\alpha$ and $\beta$ spins:
\begin{equation}
\varphi_{p}^{\alpha} \left(\mathbf{r}\right) = \varphi_{p}^{\beta} \left(\mathbf{r}\right) = \varphi_{p} \left(\mathbf{r} \right), \quad n_{p}^{\alpha}=n_{p}^{\beta}=n_{p}.
\end{equation}
It is important to note that our NOFs preserve the total spin \cite{Piris2019} regardless the value of $S$. Electron pair models require to divide the orbital space $\Omega$. We divide the latter into $\mathrm{N}_\Omega = \mathrm{N}/2$ mutually disjoint subspaces $\Omega{}_{g}$, each of which contains one orbital $\left|g\right\rangle $ with $g\leq\mathrm{N}_\Omega$, and $\mathrm{N}_{g}$ orbitals $\left|p\right\rangle $ with $p>\mathrm{N}_\Omega$, namely,
\begin{equation}
\Omega{}_{g}=\left\{ \left|g\right\rangle ,\left|p_{1}\right\rangle ,\left|p_{2}\right\rangle ,...,\left|p_{\mathrm{N}_{g}}\right\rangle \right\} .\label{OmegaG}
\end{equation}

Taking into account the spin, the total occupancy for a given subspace $\Omega{}_{g}$ is 2, which is reflected in the following sum rule: 
\begin{equation}
\sum_{p\in\Omega_g}n_{p}=n_{g}+\sum_{i=1}^{\mathrm{N}_{g}}n_{p_{i}}=1,\quad g=1,2,...,\mathrm{N}_\Omega. \label{sum1}
\end{equation}
Here, the notation $p\in\Omega_g$ represents all the
indexes of $\left|p\right\rangle $ orbitals belonging to $\Omega_g$. In general, $\mathrm{N}_{g}$ may be different for each subspace. In this work, $\mathrm{N}_{g}$ is equal to a fixed number for all subspaces $\Omega{}_{g}\in\Omega$. We adopt the maximum possible value of $\mathrm{N}_{g}$ which is determined by the number of basis functions ($N_B$) used in calculations. Indeed, variable $N_{g}$ will play a key role in the Be + H$_{2}$ reaction studied in subsection \ref{subsec:h8,beh2}. Taking into account Eq. (\ref{sum1}), the trace of the 1RDM is verified equal to the number of electrons: 
\begin{equation}
2\sum_{p\in\Omega}n_{p}=2\sum_{g=1}^{\mathrm{N}_\Omega}\sum_{p\in\Omega_g}n_{p}=2\sum_{g=1}^{\mathrm{N}_\Omega}\left(n_{g}+\sum_{i=1}^{\mathrm{N}_{g}}n_{p_{i}}\right)=\mathrm{N}\label{sumNpII}
\end{equation}

Using ensemble N-representability conditions \cite{mazziotti2012structure,S18,LCS23}, we can generate a reconstruction functional for the 2RDM in terms of the ONs that leads to GNOF \cite{piris-prl-2021}:
\begin{equation}
E=E^{intra}+E_{HF}^{inter}+E_{sta}^{inter}+E_{dyn}^{inter}
\end{equation}
For singlet states, the intra-pair component is formed by the sum of the energies of the pairs of electrons with opposite spins, namely
\begin{equation}
E^{intra}=\sum\limits _{g=1}^{\mathrm{N}_\Omega}E_{g}, \quad
E_{g} = 2 \sum\limits _{p\in\Omega_{g}}n_{p}H_{pp} + \sum\limits _{q,p\in\Omega_{g}} \Pi(n_q,n_p) L_{pq}
\end{equation}
\noindent where $\Pi(n_p,n_r) = c(n_p)c(n_r)$ and $c(n_p)$ is defined by the square root of the ONs according to the following rule:
\begin{equation}
 c(n_p) = \left. \begin{cases}
    \phantom{+}\sqrt{n_p}, & p \leq \mathrm{N}_\Omega\\
    -\sqrt{n_p}, & p > \mathrm{N}_\Omega \\
  \end{cases}  \right.,
\end{equation}
that is, the phase factor of $c_p$ is chosen to be $+1$ for a strongly occupied orbital, and $-1$ otherwise. $H_{pp}$ are the diagonal one-electron matrix elements of the kinetic energy and external potential operators and $L_{pq}=\left\langle pp|qq\right\rangle $ are the exchange-time-inversion integrals \cite{Piris1999}. The inter-pair Hartree-Fock (HF) term is
\begin{equation}
E_{HF}^{inter}=\sum\limits _{p,q=1}^{\mathrm{N}_{B}}\,'\,n_{q}n_{p}\left(2J_{pq}-K_{pq}\right)
\end{equation}
\noindent where $J_{pq}=\left\langle pq|pq\right\rangle$ and $K_{pq}=\left\langle pq|qp\right\rangle$ are the Coulomb and exchange integrals, respectively. The prime in the summation indicates that only the inter-subspace terms are taking into account ($p\in\Omega{}_{f},q\in\Omega_{g},f\neq g$). The inter-pair static component is written as
\begin{equation}
E_{sta}^{inter}=-\sum\limits _{p,q=1}^{\mathrm{N}_{B}}\,'\,
\Phi_{q}\Phi_{p} \left(1-\delta_{q\Omega^{b}}\delta_{p\Omega^{b}}\right) L_{pq}
\label{esta}
\end{equation}
\noindent where $\Phi_{p}=\sqrt{n_{p}h_{p}}$ with the hole $h_{p}=1-n_{p}$. Note that $\Phi_{p}$ has significant values only when the ON $n_p$ differs substantially from 1 and 0. In Eq. (\ref{esta}), $\Omega^{b}$ denotes the subspace composed of orbitals below the level $\mathrm{N}_\Omega$ ($p\leq\mathrm{N}_\Omega$), so interactions between orbitals belonging to $\Omega^{b}$ are excluded from $E_{sta}^{inter}$.

Finally, the inter-pair dynamic energy can be conveniently expressed as
\begin{equation}
E_{dyn}^{inter}=\sum\limits_{p,q=1}^{\mathrm{N}_{B}}\,'\,
\left[ n_{q}^{d}n_{p}^{d} + \Pi\left(n_{q}^{d},n_{p}^{d}\right) \right]
\left( 1-\delta_{q\Omega^{b}}\delta_{p\Omega^{b}} \right) L_{pq}
\end{equation}

The dynamic part of the ON is defined\cite{piris-prl-2021} as
\begin{equation}
n_{p}^{d}=n_{p}\cdot e^{-\left(\dfrac{h_{g}}{h_{c}}\right)^{2}} \label{dyn-on}
\end{equation}
\noindent with $h_{c}=0.02 \sqrt{2}$. The maximum value of $n_{p}^{d}$ is around 0.012 in accordance with the Pulay’s criterion that establishes an occupancy deviation of approximately 0.01 with respect to 1 or 0 for a NO to contribute to the dynamic correlation. Considering real spatial orbitals ($L_{pq}=K_{pq}$) and $n_{p}\approx n_{p}^{d}$, it is not difficult to verify that the terms proportional to the product of the ONs will cancel out, so that only those terms proportional to $\Pi$ will contribute significantly to the energy.

\section{Results}\label{sec:results}

The systems investigated in this study can be categorized into three groups. The first and second groups both involve the H$_4$ model, while the last group includes the symmetric dissociation of H$_8$ in one and three dimensions and the insertion of a Be atom into an H-H bond. All calculations were performed using the DoNOF code \cite{piris2021donof} with the STO-6G basis set \cite{sto-6g}, unless otherwise specified. FCI and pair Coupled Cluster Doubles (pCCD) results were obtained from Ref. \cite{gaikwad-jcp}, except for those related to the Be + H$_2$ reaction, for which the Be(3s2p)/H(2s) basis results were extracted from Ref. \cite{ammar2024gw}.

\begin{figure}[ht]
\begin{centering}
\caption{\label{planar-h4-geoms} Structure of the planar H$_{4}$ model systems studied in this work. Labels are set according to Ref. \cite{gaikwad-jcp}. \bigskip}
{\includegraphics[scale=0.3]{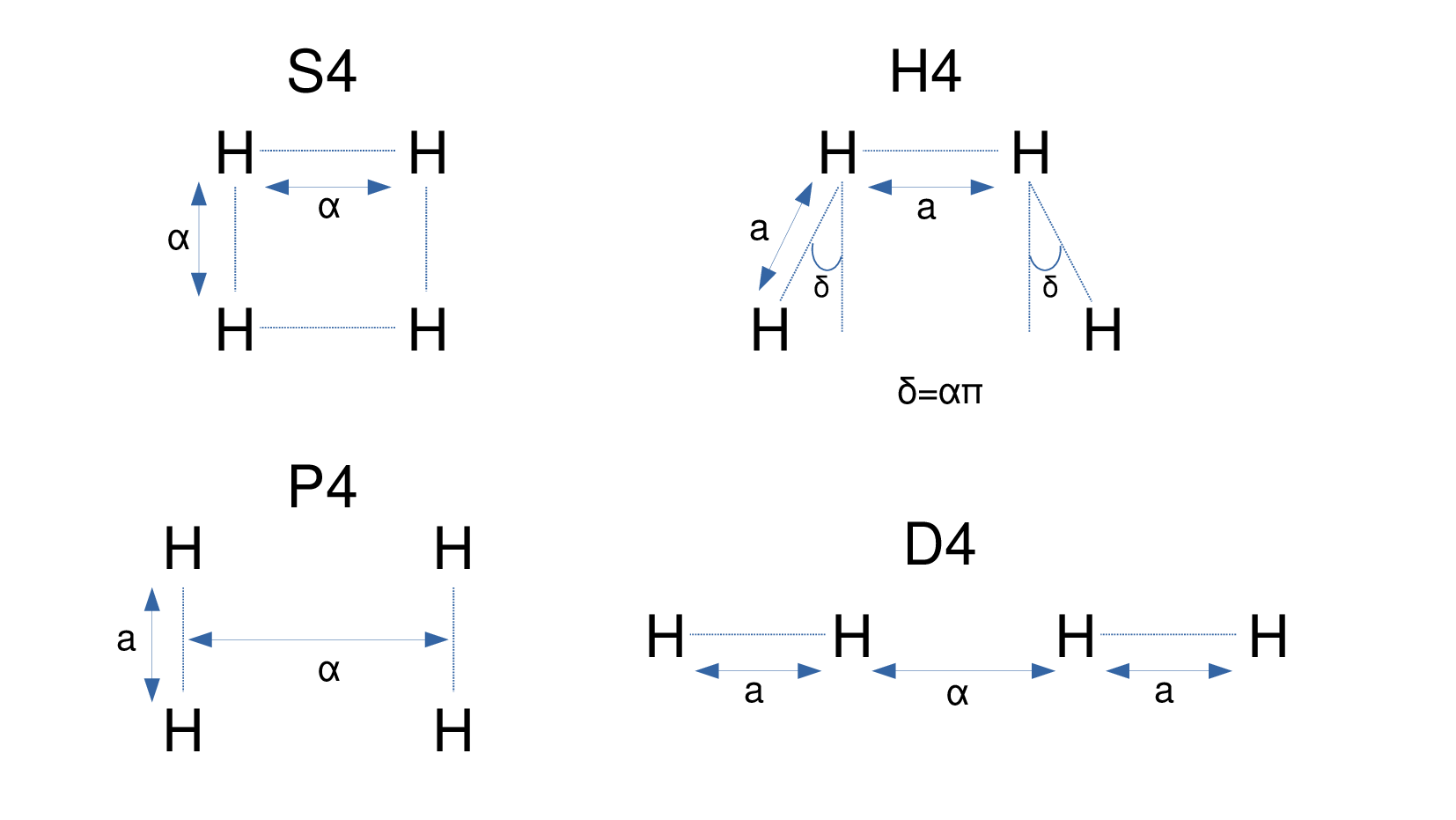}}
\end{centering}
\end{figure}

\subsection{Planar H$_{4}$ benchmark set}\label{subsec:planar}

The planar H${_4}$ system was previously investigated in the context of NOF theory by Ramos \textit{et al.} \cite{eloy-ramos-h4}. They concluded that static inter-pair correlation is crucial for accurately describing the $D_{2h}-D_{4h}$ transition. Consequently, the independent-pair model PNOF5 inaccurately exhibited a cusp along this transition. In the following sections, we examine whether GNOF can rectify the deficiencies of PNOF5 in the rectangular H${_4}$ system. Additionally, for comprehensive analysis, we include additional geometries and variants, referred to as planar systems, as depicted in Fig. \ref{planar-h4-geoms}. The notation follows that of Ref. \cite{gaikwad-jcp}, although these isomers of H${_4}$ were originally introduced by Paldus and co-workers \cite{paldus-planar-h4}.

\begin{figure}[t]
\begin{centering}
\caption{\label{planar} GNOF absolute energies in a.u.(top) and GNOF energy differences with respect to FCI in a.u.(bottom) for the planar D4, S4 and P4 systems and non-planar V4 system. Note that hydrogen bonds are fixed to $a=2.0$ Bohr in both D4 and P4, and the distance $a$ is fixed to 2.0 a.u. in V4. \bigskip}
\quad\quad {\includegraphics[scale=1.15]{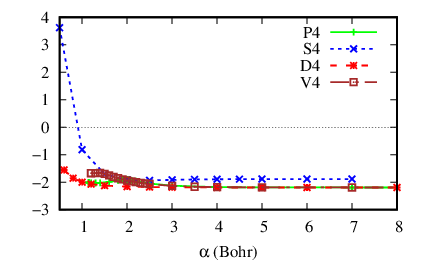}}
{\includegraphics[scale=1.2]{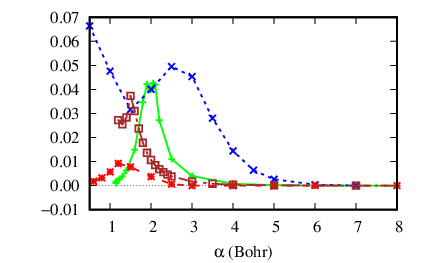}}
\end{centering}
\end{figure}

GNOF absolute energies can be observed in the top plot of Fig. \ref{planar} for the planar D4, S4, and P4 configurations, as well as V4 which is dicussed in the next section. Qualitatively, all the energy curves exhibit similar behavior: energies decrease as the distance between two hydrogen molecules increases in the case of D4 and P4, and similarly for S4, but now with respect to inter-nuclear distances. However, it is worth noting a slight shoulder in the green curve corresponding to the rectangular P4 system. This corresponds to $\alpha =2$ Bohr, the point at which P4 converges to the S4 structure.

The bottom plot in Fig. \ref{planar} illustrates GNOF energy errors for the D4, S4, and P4 model systems, together with the non-planar V4 which is studied in the following section. The former corresponds to a one-dimensional problem, where two hydrogen molecules are dissociated. Since GNOF is a quasi-exact method for two-electron systems, the error approaches zero as the correlation between two molecules decreases. According to Fig. \ref{planar}, the maximum error occurs slightly above $\alpha = 1.0 $ Bohr, where one orbital is practically doubly occupied and the other strongly occupied orbital yields an occupancy of 1.7, which represents a typical intermediate case between dynamical and static electron correlation effects. However, the maximum GNOF error remains below $0.01$ Hartree. It is noteworthy to observe the alteration in bonding patterns as D4 transitions from $\alpha < a$ to $\alpha > a$. This change is discerned through GNOF in the occupation scheme. Beyond $\alpha=2.0$ Bohr, the preference shifts towards two strongly occupied quasi-degenerate NOs and two weakly occupied quasi-degenerate NOs.

\begin{figure}[t]
\begin{centering}
\caption{\label{planar-h4} GNOF absolute energies in a.u.(top) and GNOF energy differences with respect to FCI in a.u. (bottom) for the planar H4 system with different $a$ distances (in Bohr) and the non-planar T4 system. For the latter, distance $a$ is fixed to 2.0 a.u. and $R$ values given in a.u.. \bigskip}
{\includegraphics[scale=1.1]{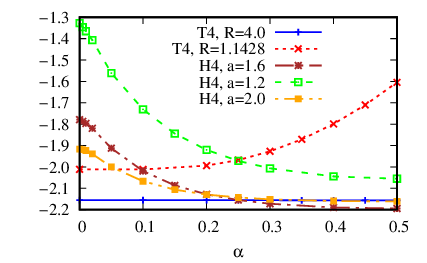}}
{\includegraphics[scale=1.15]{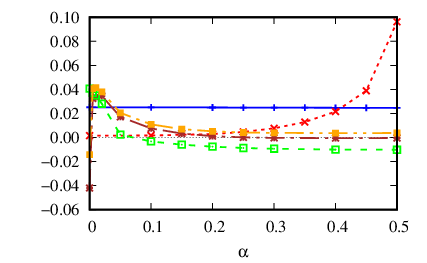}}
\end{centering}
\end{figure}

As observed in the case of the Hubbard model \cite{mitxelena2020efficient2}, the error increases as we transition from linear to two-dimensional electronic systems. In both the square and rectangular H${_4}$ systems, the agreement with FCI is remarkable beyond $a=\alpha=4.0$ Bohr. Note that both geometries converge to the same system at $a=\alpha=2.0$ Bohr. However, GNOF exhibits larger errors for the square S4 compared to the rectangular P4 geometry at both intermediate and short distances between hydrogen molecules. Interestingly, in the case of S4, GNOF yields two spatial orbitals that are singly occupied ($2n_2=2n_3=1$) along the entire dissociation curve. This may be associated with an overestimation of the static correlation at short and intermediate distances. In this system, a minimal basis set that can only provide a description of bonding-antibonding orbitals appears insufficient to capture the correlation effects adequately. Including more weakly occupied orbitals by using a larger basis set seems to be a better option, as we will demonstrate in subsection \ref{subsec:h8,beh2} for the BeH$_2$ system, where we encounter a similar problem.

On the other hand, the rectangular P4 system is more analogous to the H$_4$ model previously studied in the context of NOF approximations \cite{eloy-ramos-h4}. GNOF performs quantitatively accurately here compared to FCI at very short distances and $\alpha \geq 3.0$ Bohr. At the dissociation limit, two ONs of strongly and weakly occupied orbitals degenerate, specifically with $2n_1=2n_2=1.925$ and $2n_3=2n_4=0.075$, respectively, at $\alpha$ = 8 Bohr. However, the electronic configuration changes to four distinct doubly occupied orbitals as $\alpha$ shortens. For instance, at $\alpha$ = 1 Bohr, the occupancies are equal to 1.993, 1.980, 0.020, and 0.007. The intermediate region where both configurations compete is also well-handled by GNOF, with the error not exceeding $0.045$ Hartree. It is worth noting that P4 is a typical example wherein GNOF can find multiple solutions, making it challenging to consistently adhere to the minimum energy curve. As pointed out in Ref. \cite{johnson-deprince}, it is in this region around $\alpha$ = 2 Bohr where the dominant Slater determinant in the FCI expansion changes due to a switching of the long and short sides of the rectangle. As illustrated by Fig. \ref{figs:planar-pccd} in the Appendix B, pCCD shows larger errors than GNOF for the planar S4, P4 and D4 systems independently of the distance $\alpha$ between hydrogen molecules.

To extend our analysis of planar model systems, we investigate the trapezoidal H$_{4}$ configuration with distances between successive atoms fixed to $a=1.2$, $a=1.6$, and $a=2.0$ Bohr. As illustrated in the geometry shown in Fig. \ref{planar-h4-geoms}, when $\delta=0$, the S4 geometry is retrieved, and the D4 structure is obtained when $\delta={\pi}/{2}$. The top and bottom plots in Fig. \ref{planar-h4} display absolute and error energies corresponding to this system. Overall, the absolute energy curves resemble those obtained for the planar D4, S4, and P4 systems shown in Fig. \ref{planar}. It is important to note, however, that Fig. \ref{planar-h4} presents energies with respect to the parameter $\alpha$, which causes the structure to vary between S4 and D4 as described above. Hence, for any $a$ distance between hydrogen molecules, the minimum energy is obtained for the linear geometry and the maximum energy for the square configuration.

In comparison with the planar models studied above, the performance of GNOF remains consistent for trapezoidal systems, as evident from the difference energy curves shown in the bottom plot of Fig. \ref{planar-h4}. Johnson and DePrince pointed out \cite{johnson-deprince} that trapezoidal H$_4$ systems require a description of weak correlation effects not present in the aforementioned S4 and P4 systems. Despite posing a challenge for the orbital-optimized doubly-occupied configuration interaction method, these systems are well-described by GNOF. In fact, a comparison with pCCD reveals that the error of GNOF is lower than that of pCCD for all H4 geometries (see Fig. \ref{figs:planar-h4-pccd} in Appendix B), so the improvement of GNOF over pCCD is more significant when $a$ increases.

We observe small deviations in energy below FCI at $\alpha \geq 0.1$ for the $a=1.2$ Bohr case (bottom plot of Fig. \ref{planar}). Although details regarding N-representability are omitted in Section \ref{sec:theory}, it is worth noting that the reconstructed 2RDM that leads to GNOF only satisfies some necessary N-representability conditions. Consequently, violations of functional N-representability \cite{Ludena2013,Piris2018-nrep} may occur, leading GNOF to result in energies below FCI values. On the other hand, the 1RDM is an ensemble N-representable matrix since the ONs belong to the interval [0,1]. Additionally, it also satisfies the pairing conditions (\ref{sum1}) linked to pure N-representability \cite{Piris2018-nrep}, allowing the avoidance of obtaining fractional charges in homolytic dissociations \cite{Matxain2011,Lopez2012}.

\subsection{Non-Planar H$_{4}$ benchmark set}\label{subsec:nonplanar}

\begin{figure}[ht]
\begin{centering}
\caption{\label{nonplanar-h4-geoms} Structure of the non-planar H$_{4}$ model systems studied in this work. Labels set according to Ref. \cite{gaikwad-jcp}. \bigskip}
{\includegraphics[scale=0.3]{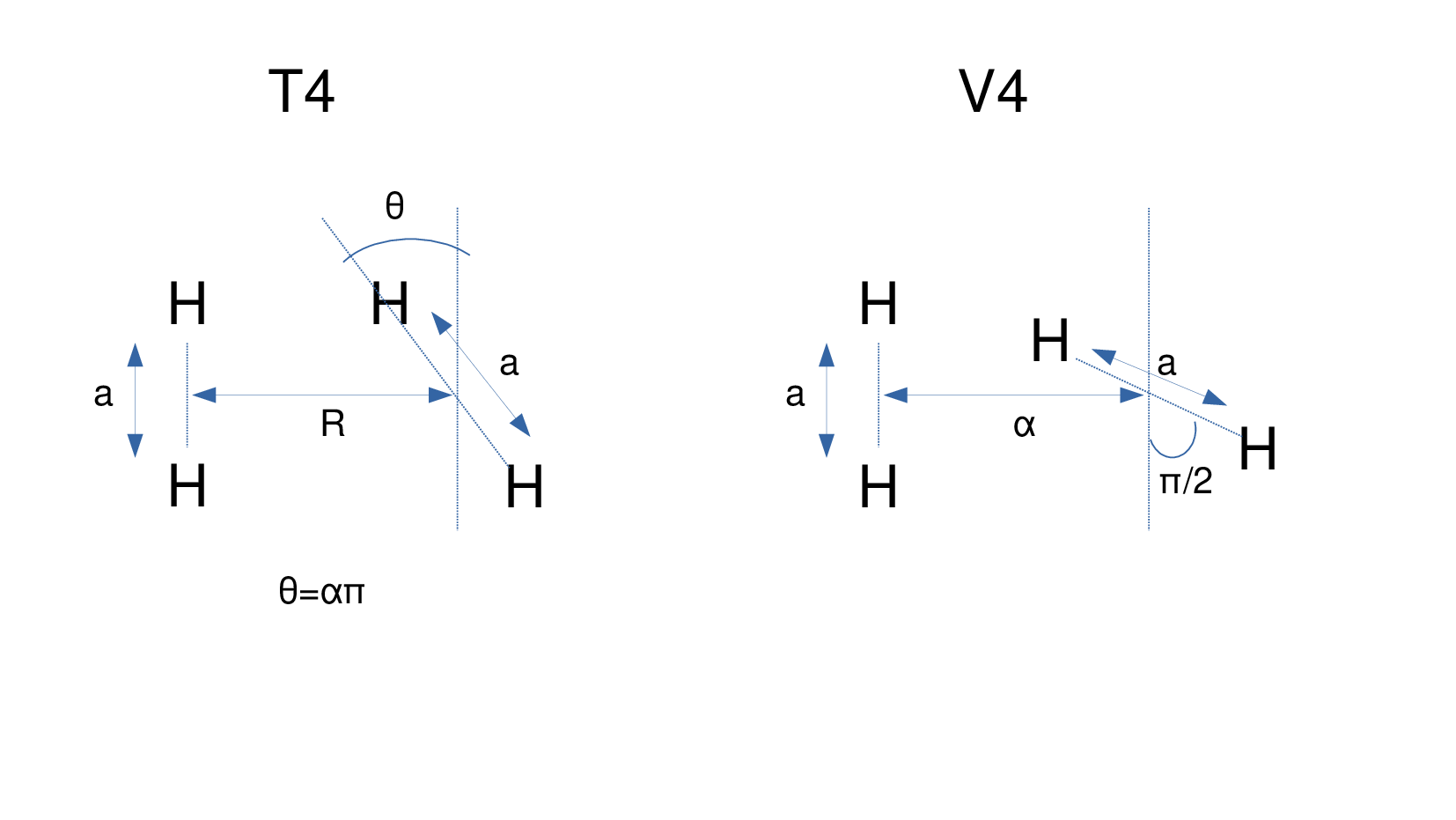}}
\end{centering}
\end{figure}

The amount of correlation energy is strongly related to the dimensionality of electronic systems. This phenomenon was demonstrated in the context of NOF approximations by studying the Hubbard model in one \cite{mitxelena2020efficient1} and two dimensions \cite{mitxelena2020efficient2}. Here, we extend the benchmark conducted in the previous section to three-dimensional H$_4$ configurations. The geometries involved are illustrated in Fig. \ref{nonplanar-h4-geoms}, where the notation is again taken from Ref. \cite{gaikwad-jcp}.

As shown in Fig. \ref{nonplanar-h4-geoms}, the H$_4$ systems studied in this section can be obtained by twisting the rectangular P4 model shown in the previous section. In the case of T4, the angle $\theta$ varies in the range $[0,\pi/2]$ for many fixed distances between the two centers of hydrogen molecules, whereas for V4, the angle between two bonds is set to $\pi/2$ and it is the distance which increases from $\alpha = 1.0$ to $\alpha = 10.0$ Bohr. For the latter, GNOF gives energies similar to the rectangular P4 shown in the bottom panel of Fig. \ref{planar}. Brown curve in Fig. \ref{planar} shows GNOF absolute energies and errors with respect to FCI for V4 in terms of the distance between two hydrogens specified above. While at intermediate distance ($\alpha \approx 1.5$ Bohr) the error approaches $0.04$ Hartree, it rapidly decreases at both shorter and longer distances. In comparison with pCCD, Fig. \ref{figs:non-planar-pccd-v4} in Appendix C shows an improvement of GNOF over pCCD close to $0.06$ and $0.04$ Hartree at long and short $\alpha$ distances, respectively.
As in many other cases, the orbital scheme is different in these regions. GNOF gives two singly occupied spatial orbitals with half ONs ($2n_3=2n_4 \approx 1.0$) at short distances, but then it changes smoothly to an HF-like configuration with two markedly occupied orbitals and corresponding weakly occupied orbitals at long distances. Regarding absolute energies, the brown curve in the top plot of Fig. \ref{planar} seems similar to the error since, after a slight increase in the energy from $\alpha \approx 1.1$ to $\alpha \approx 1.5$ Bohr, it rapidly decreases to a constant beyond $\alpha \approx 4$ Bohr.

As illustrated by blue and red curves in Fig. \ref{planar-h4}, for T4, the situation changes dramatically depending on the fixed distance between two hydrogen bonds. When the two hydrogen molecules are brought as close as $R=1.1428$ Bohr, the GNOF error stays under $0.04$ Hartree for small angles up to approximately $0.45\pi$. However, the performance rapidly deteriorates from this point to $\theta = \pi/2$, where we obtain a difference $E_{\text{GNOF}}-E_{\text{FCI}}$ close to $0.1$ Hartree. According to the top plot of Fig. \ref{planar-h4} corresponding to absolute energies, as the energy increases and the red curve ascends, the error with respect to FCI also increases, indicating that GNOF is not able to capture a significant portion of dynamic electron correlation in this region. Natural ONs are close to integer values along the full red curve, so a larger basis set allowing a larger $\mathrm{N}_{g}$ value is necessary to improve GNOF performance, as happened for BeH$_2$ in subsection \ref{subsec:h8,beh2}.

Similar to the behavior shown by geminal wavefunctions developed by Gaikwad and co-workers in Ref. \cite{gaikwad-jcp}, after reaching some distance close to $R =2$ Bohr, the difference with respect to FCI remains nearly constant for any angle between two hydrogen bonds, as illustrated by the blue solid curve in Fig. \ref{planar-h4} corresponding to $R=4$ Bohr. There are no significant changes in the ONs and NOs when $\theta$ varies between $0$ and $\pi/2$, so GNOF exhibits two quasi-degenerate orbitals with ONs close to $0.98$ and corresponding weakly occupied orbitals with ONs $\approx 0.005$. This suggests that the electronic problem is dominated by dynamic correlation.

Results corresponding to the T4 geometry with distances $R = \sqrt{2}, 2, 3, 7$ a.u. can be found in the Appendix C. The latter also includes Fig. \ref{figs:non-planar-pccd-t4} to compare GNOF and pCCD performances on the non-planar T4 system with $R = 1.1428, 4$ a.u. and $a = 2.0$ a.u. While GNOF shows a lower error for any $\alpha$ value when $R = 4$ a.u., there is one point corresponding to $\alpha = 0.5$ where pCCD improves the GNOF error when $R = 1.1428$ a.u.. Overall, GNOF error is $0.02$ a.u. lower than pCCD.

\subsection{Symmetric dissociation of linear and cubic H$_{8}$ and the Be+H$_{2}$ reaction}\label{subsec:h8,beh2}

\begin{figure}[ht]
\begin{centering}
\caption{\label{h8-geoms} Structures corresponding to the linear and cubic H$_{8}$, together with the structure corresponding to the Be+H$_2$ reaction based on Fig. 1 from Ref. \cite{ammar2024gw}. Note beryllium is placed at the coordinate origin, whereas hydrogens are at $\pm y=2.54-0.46x$ varying $x$ from $0$ to $4$ Bohr.. \bigskip}
{\includegraphics[scale=0.28]{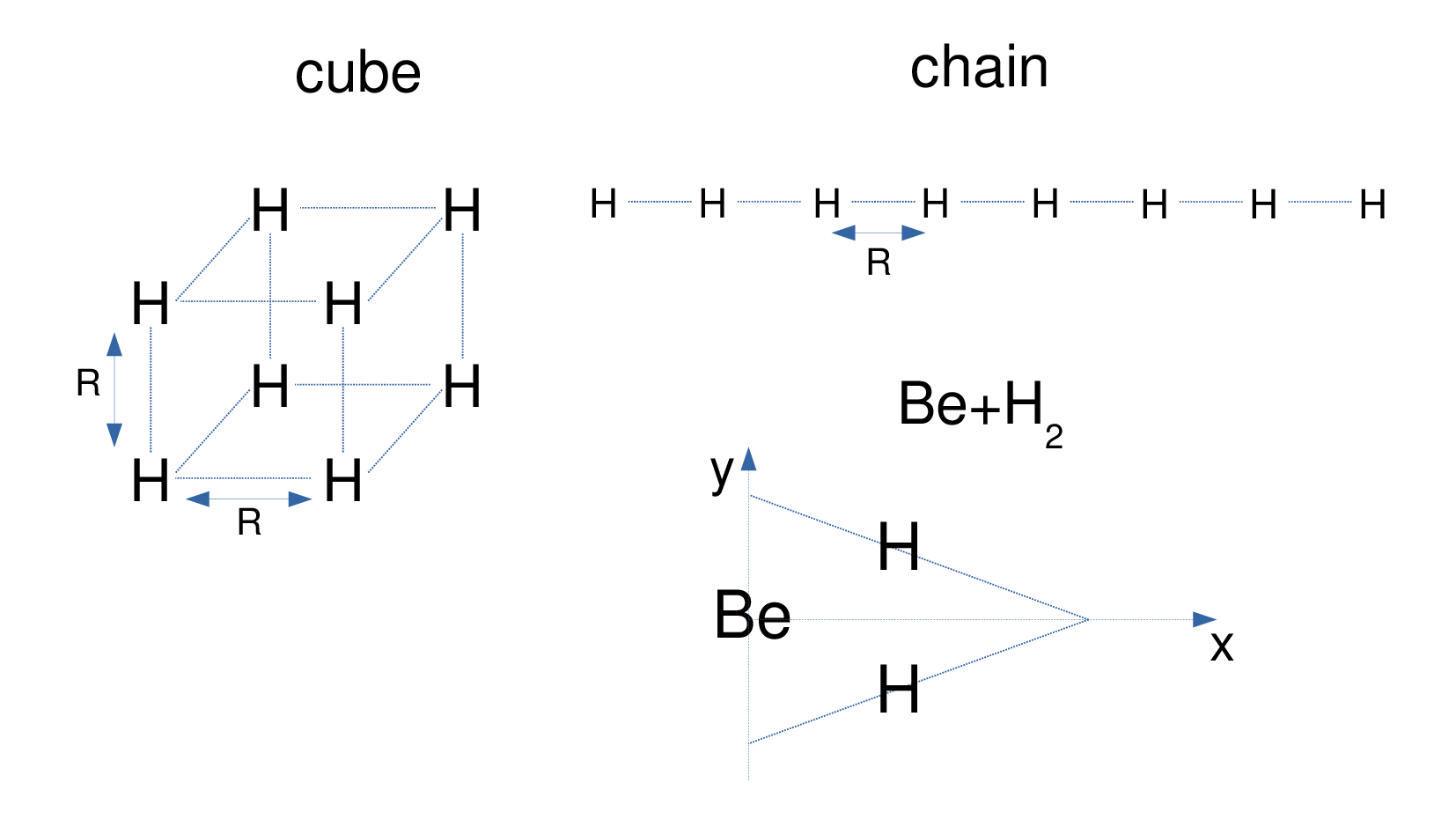}}
\end{centering}
\end{figure}

Following the work by Gaikwad \textit{et al.} \cite{gaikwad-jcp}, we conclude the assessment of GNOF with three paradigmatic cases: the dissociation of a 3D H${_8}$ cube, a linear chain of eight hydrogen atoms, and the insertion of a beryllium atom into a hydrogen molecule. Similar systems to the former were studied in Ref. \citep{mitxelena_2022}. Despite the qualitatively correct dissociation curve provided by GNOF in Ref. \cite{mitxelena_2022} for H$_{64}$ in three dimensions, the lack of a reliable reference method motivates us to include a similar system such as the cubic H${_8}$ in the present work. Geometries and notation corresponding to the H${_8}$ systems are shown in Fig. \ref{h8-geoms}. Note that symmetric dissociation refers to stretching all bonds simultaneously to obtain, in the end, 8 isolated hydrogen atoms. When the correlation regime is increased, a metal-to-insulator transition is observed in both systems. Accordingly, they are paradigmatic models for strongly correlated Mott insulators.

\begin{figure}[htbp]
\caption{\label{h8-cube} GNOF, pCCD and FCI absolute energies in a.u. (top) and GNOF and pCCD energy differences with respect to FCI in a.u. (bottom) for the linear and cubic H$_{8}$. \bigskip}
\includegraphics[scale=1.2]{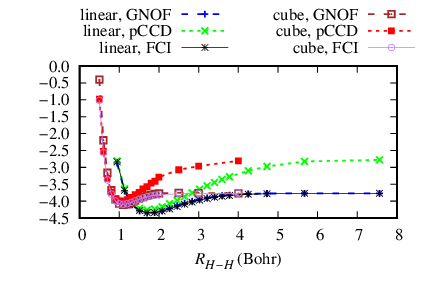}
\includegraphics[scale=1.2]{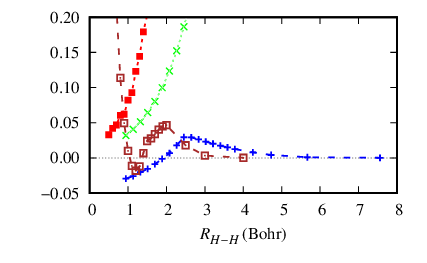}
\end{figure}

Fig. \ref{h8-cube} shows absolute energies and energy differences with respect to FCI given by GNOF and pCCD for the cubic and linear H${_8}$ systems. As expected, at large distances in the strong correlation regime, the error rapidly approaches zero in both model systems. However, as illustrated by the brown curve at the bottom plot of Fig. \ref{h8-cube}, the performance at intermediate and short distances breaks down in the three-dimensional case. Let us focus on the latter, since here the error increases up to one order of magnitude larger than at intermediate distances. This system resembles the S4 model studied in Fig. \ref{planar}. Unfortunately, in these systems, GNOF struggles to retrieve dynamic correlation at very short distances with a minimal basis set, so it is necessary to include more weakly occupied orbitals to improve GNOF energies in this region. Regarding the GNOF and pCCD comparison, pCCD gives accurate energies at short inter-nuclear distances in both linear and cubic H${_8}$, where dynamic correlation effects are dominant. However, pCCD error increases dramatically with the internuclear distance and it cannot describe the electron localization.

Regarding H${_8}$ dissociation in one dimension, blue curves in Fig. \ref{h8-cube} confirms the conclusions obtained in Ref. \cite{mitxelena2020efficient1} for GNOF predecessors. At large distances, GNOF retrieves FCI energies. However, GNOF misses some correlation energy at intermediate H-H distances; for instance, there is a difference of approximately $0.03$ Hartree at $R{_\text{H-H}}=2.5$ a.u., where two strongly occupied orbitals yield $2n_{1}=2n_{2} \approx 1.80$. It is worth noting that GNOF goes below FCI when hydrogen atoms are brought very close to each other. Overall, plots corresponding to absolute energies show qualitatively good behavior for both systems and resemble typical dissociation curves in molecular systems. A comparison between absolute energy and energy error plots reveals that while equilibrium and dissociation limit regions are accurately described by GNOF, intermediate and short bond distance regions are especially challenging.

\begin{figure}[t]
\begin{centering}
\caption{\label{beh2} GNOF, pCCD and FCI absolute energies in a.u. (top) and GNOF and pCCD energy differences with respect to FCI in a.u. (bottom) for Be+H$_{2}$ reaction. Note legend to distinguish between STO-6G and Be(3s2p)/H(2s) basis sets, as well as $N_{g}$ values and different methods. \bigskip}
{\includegraphics[scale=1.0]{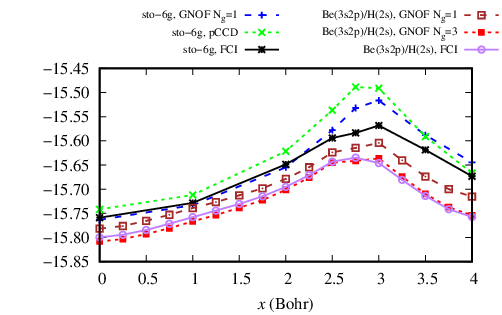}}
{\includegraphics[scale=1.2]{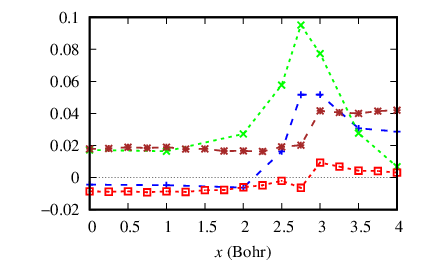}}
\end{centering}
\end{figure}

BeH$_2$ is employed here to represent a chemical reaction involving strong correlation. Recently, Ammar and co-workers \cite{ammar2024gw} have studied the ability of the GW approximation to describe the strong correlation region of the Be+H$_2$ reaction. In fact, when a Be atom is inserted into H$_2$ following a C${_{2v}}$ pathway as illustrated in Fig. \ref{h8-geoms}, the FCI wavefunction of the whole system is dominated by two different electronic configurations depending on the distance between beryllium and H$_2$ center. Hence, the switching region is plagued by strong correlation effects. It is worth noting that different basis sets were employed in Refs. \cite{ammar2024gw} and \cite{gaikwad-jcp} to study the strong correlation of the Be+H$_2$ reaction. While STO-6G was used in the latter, the former utilized the Be(3s2p)/H(2s) configuration. In Fig. \ref{beh2}, we show GNOF and pCCD absolute energies and energy differences with respect to FCI along the beryllium insertion into H$_2$, respectively in the top and bottom panels. It is worth noting that GNOF error remains below pCCD for any $x$ distance. The difference is significant at $2.5a_{0}< x < 3a_{0}$, a region which is discussed in detail below.

In Ref. \cite{ammar2024gw}, Ammar \textit{et al.} point out that in the region $2.5a_{0}< x < 3a_{0}$, the FCI wavefunction switches from one electronic configuration to another. Therefore, it is at $x=2.75a_{0}$ where both dominant configurations have the same weight, and maximum strong correlation effects might occur. Looking at absolute energies in the top plot of Fig. \ref{beh2}, there is an increase in energy in this region, which is significantly weaker if the basis set is large enough, as revealed by the red curve in comparison with the green and blue curves. In other words, GNOF has no problem describing this region as long as the orbital subspace containing an electron pair is large enough, as revealed by the energy errors in the bottom panel of Fig. \ref{beh2}. This issue is evident in the bottom plot of Fig. \ref{beh2} when we focus on the red curve, which keeps the difference with respect to FCI below 0.01 hartree.

It is worth recalling here the meaning of the parameter $\mathrm{N}_{g}=1$, which corresponds to a GNOF calculation with a minimal basis set, i.e., two orbitals per electron pair. According to the description given in section \ref{sec:theory}, this means having one strongly occupied orbital ($n_p \geq 0.5$) coupled with a weakly occupied one ($n_p \leq 0.5$). Looking at the ONs obtained at $x=2.75a_{0}$, we have two orbitals almost fully occupied ($n_{1} \approx 0.9998$, $n_{2} \approx 0.9825$), two orbitals half-occupied ($n_{3} \approx 0.5$, $n_{4} \approx 0.5$), and two almost empty orbitals ($n_{5} \approx 0.035$, $n_{6} \approx 0.0004$). Nevertheless, when the orbital subspace is extended to four orbitals and thereby each strongly occupied NO is coupled with three weakly occupied NOs ($\mathrm{N}_{g}=3$), we have a similar distribution of the ONs, except for the weakly occupied orbitals coupled to orbital number 2, which present a non-negligible distribution of the ONs. This makes the difference with respect to FCI nearly zero. More details about the ONs corresponding to this system can be found in the Appendix A.

\bigskip
\section{Conclusions}\label{sec:conclusions}

This paper examines the performance of the Global Natural Orbital Functional (GNOF) approximation when applied to hydrogen atom clusters in one, two, and three dimensions, comparing its outcomes with those obtained from the FCI method. The GNOF energy expression, relying solely on two-index electron integrals, minimizes computational scaling with system size while maintaining the capacity to describe diverse types of electron correlation. This feature renders it highly attractive as an approximation method.

In this study, GNOF has demonstrated its effectiveness in capturing strong correlations across various scenarios. Three distinct blocks were distinguished, with the first and second corresponding to known groups of H$_4$ atoms. In these cases, GNOF errors remain below $0.05$ Hartree, with the intermediate region showing maximum differences when dynamic and static electron correlation effects compete. Notably, challenges arose in the T$_4$ system, where two hydrogen molecules rotate with respect to each other, particularly when the distance between H$_2$ is small.

The third block focused on more intricate systems, including the symmetric dissociation of two H$_8$ systems. In one dimension, GNOF yielded some energies below those of FCI at very short internuclear distances, yet the overall accuracy remained high. However, during the dissociation of the H$_8$ cube, significant deviations between GNOF and FCI were observed. While agreement with FCI was accurate at bond H-H distances exceeding 1 Bohr, the error rapidly increased to approximately $0.55$ Hartree as the internuclear H-H distance decreased to 0.5 Bohr. The assessment culminated with a typical reaction demonstrating strong correlation effects: the insertion of a beryllium atom into a hydrogen molecule. This reaction underscored the importance of expanding the number of orbitals within each orbital subspace. Indeed, GNOF achieved greater precision with an augmented basis set, transitioning from STO-6G to Be(3s2p)/H(2s). 

In summary, the challenges encountered by GNOF in the T4 system, the dissociation of the H$_8$ cube, and the Be+H$_2$ reaction underscore the difficulty of capturing dynamical correlation at extremely short distances using minimal basis sets. This highlights the necessity of incorporating more weakly occupied orbitals to enhance GNOF energies in such scenarios. Coupled with recent publications endorsing the efficacy of GNOF across various atomic and molecular systems, this study reinforces GNOF as a well-rounded method for addressing global electronic structure problems.

\bigskip
\begin{acknowledgments}
Financial support comes from the Eusko Jaurlaritza (Basque Government), Ref.: IT1584-22 and from the Grant No. PID 2021-126714NB-I00, funded by MCIN/AEI/10.13039/501100011033. The authors also the technical and human support provided by the IZO-SGI SGIker of UPV/EHU and DIPC for the generous allocation of computational resources. 
\end{acknowledgments}

\appendix

\section{GNOF occupation numbers for BeH$_{2}$}
\begin{table}[ht!]
  \caption{ONs for BeH$_{2}$ at $x=3$ Bohr for $\mathrm{N}_{g}=1$ and $\mathrm{N}_{g}=3$.}\bigskip
  \begin{center}
  \begin{tabular}{ccc}    \hline
$n$	&$\mathrm{N}_{g}=1$    &$\mathrm{N}_{g}=3$ \\ \hline
1 & 0.99970 & 0.99922 \\
2 & 0.98533 & 0.97941 \\
3 & 0.50000 & 0.52503 \\
4 & 0.50000 & 0.47462 \\
5 & 0.01467 & 0.01027 \\
6 & 0.00030 & 0.00723 \\
7 & 0.00000 & 0.00308 \\
8 & 0.00000 & 0.00027 \\
9 & 0.00000 & 0.00027 \\
10 & 0.00000 & 0.00026 \\
11 & 0.00000 & 0.00025 \\
12 & 0.00000 & 0.00008 \\
13 & 0.00000 & 0.00000 \\ \hline
    \end{tabular}
    \end{center}
\end{table}

\section{Planar H$_{4}$ benchmark set}
\begin{figure*}[ht!]
  \includegraphics[scale=1.8]{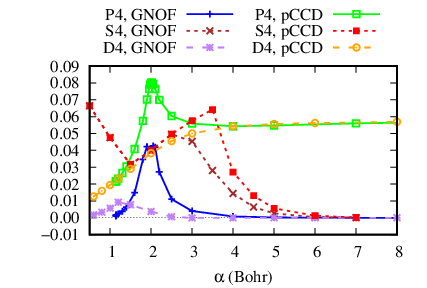}
\caption{GNOF and pCCD energy differences with respect to FCI in a.u. for the planar D4, S4 and P4 systems. Note that hydrogen bonds are fixed to $a=2.0$ Bohr in both D4 and P4.} 
\bigskip
\label{figs:planar-pccd}
\end{figure*}
\begin{figure*}[ht!]
  \includegraphics[scale=1.8]{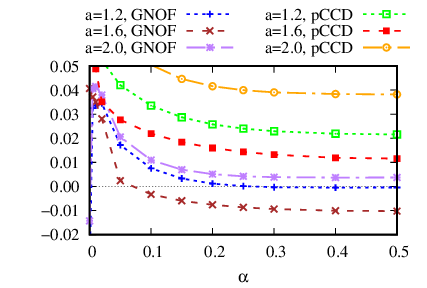}
\caption{GNOF and pCCD energy differences with respect to FCI in a.u. for the planar H4 systems.} 
\bigskip
\label{figs:planar-h4-pccd}
\end{figure*}

\section{Non-Planar H$_{4}$ benchmark set}
\begin{figure*}[ht!]
  \includegraphics[scale=1.0]{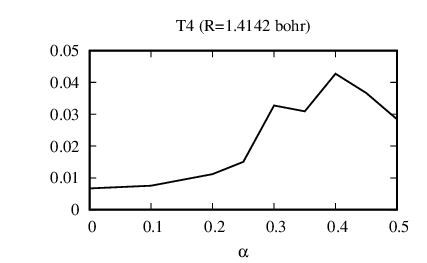}
  \includegraphics[scale=1.0]{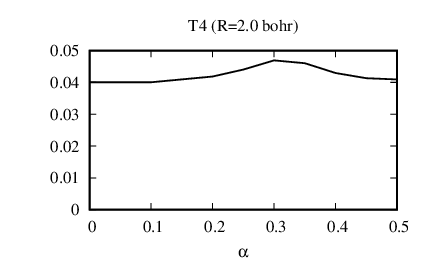}
  \includegraphics[scale=1.0]{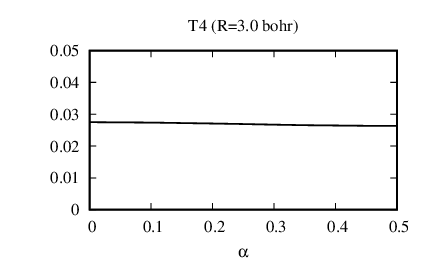}
  \includegraphics[scale=1.0]{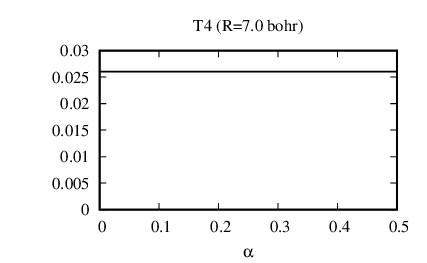}
\caption{GNOF energy differences with respect to FCI (in a.u.) for the non-planar T4 system at $R=\sqrt{2}$ (top-left) $R=2$ (top-right) $R=3$ (bottom-left), and $R=7$ Bohr (bottom-right).} 
\bigskip
\label{figs:non-planar-t4}
\end{figure*}

\begin{figure*}[h!]
  \includegraphics[scale=1.8]{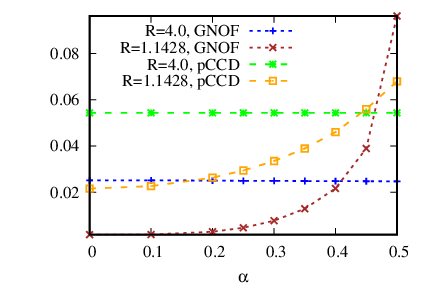}
\caption{GNOF and pCCD energy differences with respect to FCI in a.u. for the non-planar T4 system with $a = 2.0$ a.u. and $R = 1.1428, 4$ a.u..} 
\bigskip
\label{figs:non-planar-pccd-t4}
\end{figure*}
\begin{figure*}[h!]
  \includegraphics[scale=1.8]{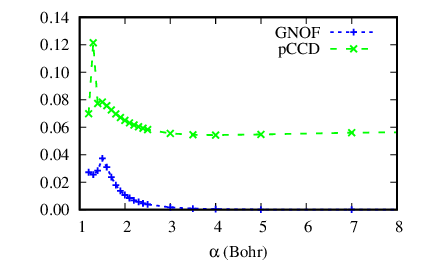}
\caption{GNOF and pCCD energy differences with respect to FCI in a.u. for the non-planar V4 system with $a = 2.0$ a.u..}
\bigskip
\label{figs:non-planar-pccd-v4}
\end{figure*}

\end{document}